\documentclass[preprint]{elsarticle}

\usepackage{etex,amssymb,amsthm,amscd,amsbsy,array}

\usepackage{lineno,hyperref}

\modulolinenumbers[5]

\journal{Memoriam Raymond Stora, NPB}

\newfont{\goth}{eufm10 scaled\magstep1}
\newfont{\tenmsb}{msbm10 scaled\magstep1}

\let\ssection=\section
\renewcommand{\section}{\setcounter{equation}{0}\ssection}

\newcommand{\medbox}[1]{\fbox{%
\rule[-10pt]{0pt}{25pt}$\;\;\displaystyle{#1}\;\;$}%
}

\newcommand{\gothg}{\mbox{\goth g}}

\newcommand{\half}{{\scriptstyle{\frac{1}{2}}}}

\newcommand{\cE}{{\cal E}}

\newcommand{\vcJ}{\mathbf{J}}

\newcommand{\cM}{{\cal M}}

\newcommand{\bR}{{\bf R}}

\newcommand{\barI}{\overline{I}}
\newcommand{\barJ}{\overline{J}}
\newcommand{\barP}{\overline{P}}
\newcommand{\barU}{\overline{U}}
\newcommand{\barX}{\overline{X}}

\newcommand{\hsigma}{\sigma_{\mathrm{RS}}}

\def\Tr{\mathop{\rm Tr}\nolimits}

\newcommand{\rg}{\mathrm{g}}

\def\SO{\mathop{\rm SO}\nolimits}
\def\so{\mathop{\rm so}\nolimits}
\def\SE{\mathop{\rm SE}\nolimits}

\def\vE{{\mathbf{E}}}

\def\vL{{\mathbf{L}}}
\def\vp{{\mathbf{p}}}
\def\vr{{\mathbf{r}}}
\def\vu{{\mathbf{u}}}
\def\vv{{\mathbf{v}}}

\def\v0{{\mathbf{0}}}

\def\vS{{\mathbf{S}}}

\newcommand{\vomega}{\boldsymbol{\omega}}

\newcommand{\la}{{\langle}}
\newcommand{\ra}{{\rangle}}

\newcommand{\uv}{{\la\vu,\vv\ra}}

\newcommand{\where}{\hbox{where}}
\newcommand{\with}{\hbox{with}}

\begin{document}

\begin{frontmatter}

\title{On the Bargmann-Michel-Telegdi equations, and\\
spin-orbit coupling:
a tribute to Raymond Stora}

\author{Christian Duval\fnref{myfootnote}}
\address{Centre de Physique Th\'eorique, Aix Marseille Universit\'e \& Universit\'e de Toulon\\ \& CNRS UMR 7332}
\ead{duval@cpt.univ-mrs.fr}
\ead[url]{www.cpt.univ-mrs.fr}




\address[mymainaddress]{Case 907, Luminy Campus, 13288 Marseille, Cedex 09 (France)}

\begin{abstract}
The Bargmann-Michel-Telegdi equations describing the motions of a spinning, charged, relativistic particle endowed with an anomalous magnetic moment in an  electromagnetic field, are reconsidered. They are shown to duly stem from the linearization of the characteristic distribution of a pre\-symplectic structure refining the original one of Souriau. In this model, once specialized to the case of a static electric-like field, the angular momentum and energy given by the associated moment map now correctly restore the spin-orbit coupling term. This is the state-of-the-art of unfinished joint work with Raymond Stora.
\end{abstract}


\end{frontmatter}

\section{Forewords}\label{ForewordsSection}

Over a decade ago, I have been informed by Serge Lazzarini that Raymond Stora and Valentine Telegdi were discussing, at CERN, an issue related to the expression of the spin-orbit coupling term in Souriau's \textit{classical} pre\-symplectic model of spinning particles leading to the Bargmann-Michel-Telegdi (BMT) equations. Stora chose that model owing of the lack of a proper Lagrangian formal\-ism for these equations. A number of stimulating exchanges with him then convinced me that the Souriau model (recal\-led in Section~\ref{SectionJMS}) should be somehow re\-visited so as to yield again the ``robust'' BMT equations as well as the correct expression of the spin-orbit term, effectively model-dependent in the considered classical framework, and which was missing in the above-mentioned model. In his carefully hand-written notes, Stora put forward an ingenious Ansatz, presented in Section~\ref{SectionRS}, which proved quite useful to meet the above-mentioned requirements. We discussed the merits and demerits of this new model which features two, a priori independent, phenomeno\-logical para\-meters in the definition (\ref{RS2form}) of the presymplectic (Lagrange) two-form~$\hsigma$. 
This flexibility finally enabled us to determine these adjustable parameters, hence the sought after model that would guarantee the BMT equations (in the weak field limit) with gyromagnetic ratio $g$, and also provide the standard spin-orbit coefficient, proportional to $g-1$, usually deduced from the Dirac-Pauli equation in the quantum mechanical framework. This new cons\-truction is patently pheno\-menological, as is Souriau's one yielding a coefficient $g$, instead of $g-1$. Now, the obtained fixing (\ref{goodCoeffs}) of the coefficients of our model should have preferably resulted from another, fully ``predictive'' model. This difficulty should be imputed to the fundamental difference between the physics ruled by quantum mechanics and that described by (semi-)classical models, even by those with a sound geometric basis. The arduous quest of this conjectured predictive model has, since then, been placed on standby. After Raymond's  passing, I found it fair and useful to make accessible to our com\-munity one of his yet unanswered queries and to witness his great scientific insights.

\goodbreak

\section{The BMT equations}\label{BMTsection}

Consider a relativistic particle with mass $m$, spin $s$, electric charge $q$, and 
\textit{gyro\-magnetic} ratio $g$, under the influence of an external and constant electro\-magnetic field $F$ in Minkowski space-time $\bR^{3,1}$, with metric $\mathrm{g}$ of
signature $-3+1$.

Let $P$ and $S$ denote respectively the linear momentum and 
spin vectors  of our particle at  space-time location $X$. The celebrated Bargmann-Michel-Telegdi equations governing the classical motions of the particle with electric charge, $q$, and gyromagnetic ratio, $g$, read \cite{BMT}
\begin{equation}
\left\{
\begin{array}{rcl}
\displaystyle
\frac{dX}{d\tau}&=&P\\[10pt]
\displaystyle
\frac{dP}{d\tau}&=&\displaystyle
-q\,FP\\[10pt]
\displaystyle
\frac{dS}{d\tau}&=&\displaystyle
-q\left[\frac{g}{2}\,FS + \left(1-\frac{g}{2}\right)\frac{P\,(\barP{}FS)}{P^2}
\right]\\
\end{array}
\right.
\label{BMT}
\end{equation}
where $\tau$ is proportional to proper time.\footnote{The bar in the notation $\barP=\rg(P,\cdot\,)$ denotes $\rg$-transposition, and $P^2=\barP{}P$ is used as a shorthand. The two-forms, $F$, are often traded as fields of skewsymmetric endomorphisms of space-time, $F=-\overline{F}$, via $F(P,S)\equiv\barP{}FS$. Space-time is oriented, and time-orientation is also assumed. We will put $c=1$ throughout the paper.}

These equations describe the motions of the particle in a \textit{weak}
electro\-magnetic field. Let us emphasize that the spin precession featured by the third equation in (\ref{BMT})
is used by experimentalists to measure $g-2$ with a very high accuracy; see, e.g., \cite{JN} for a survey of the field-theoretical computations and experimental data concerning the muon gyromagnetic ratio, $g_\mu$.
We recall that the BMT equations have first been deduced from a semi-classical approximation of the Dirac-Pauli equation in \cite{RK}. 

\goodbreak


\section{Spinning particles in an electromagnetic field: Souriau's model}
\label{SectionJMS}

The system (\ref{BMT}) was reconsidered by Jean-Marie Souriau (JMS) in the
framework of presymplectic mechanics associated, in the relativistic framework,  with the classification of co\-adjoint orbits of the (connected component of the) Poincar\'e group, $\SE(3,1)$, i.e., of elementary relativistic classical systems \cite{SSD}.

\subsection{The free case}

It has been shown that the BMT equations can be given by the kernel of a certain
presymplectic two-form on the nine-dimensional \textit{evolution space} 
\begin{equation} 
V=\{(X,I,J)\in{}(\bR^{3,1})^3\,\big\vert\,I^2=-J^2=1,\barI{}J=0, I\ \hbox{future-pointing}\}
\label{V}
\end{equation} 
which is diffeomorphic to $\SE_+(3,1)/\SO(2)$.

Before dwelling on the coupling to an external electromagnetic field, let us first recall how to describe the motions of a free relativistic particle endowed with mass, $m$, and spin, $s$. Start with the closed two-form on~$V$ given by \cite{SSD}
\begin{equation}
\sigma_\mathrm{JMS}^\mathrm{free}
=
\half{}s\left(d\barI\wedge\Omega\,dI-d\barJ\wedge\Omega\,dJ\right)
-m\,d\barI\wedge{}dX
\label{free2form}
\end{equation}
where
\begin{equation}
\Omega=j(I,J)
\label{Omega}
\end{equation}
is the normalized spin tensor, i.e., the $\so(3,1)$-valued cross-product of $I$ and~$J$.\footnote{One has
$\Omega(I,J)=\star(I\barJ-J\barI)$, where ``$\star$'' stands for the Hodge-star.
} 
The unparametrized equations of free motion, namely $\delta(X,I,J)=(\lambda{}I,0,0)$, with $\lambda\in\bR$, are given by the null distribution of (\ref{free2form}). As to the manifold $V/\ker(\sigma_\mathrm{JMS}^\mathrm{free})$ of all the solutions of these equations, it is actually symplectomorphic to the eight-dimensional $\SE_+(3,1)$-coadjoint orbit characterized by the invariants $m$ and $s$.

\subsection{Electromagnetic coupling}

In order to switch on non-minimal electromagnetic coupling of the system to an otherwise arbitrary external electromagnetic field (EMF), $F$,
Souriau has proposed to replace (\ref{free2form}) by the presymplectic (Lagrange) two-form \cite{SSD,JMS}
\begin{equation}
\sigma_\mathrm{JMS}
=
\half{}s\left(d\barI\wedge\Omega\,dI-d\barJ\wedge\Omega\,dJ\right)
-d(M\barI)\wedge{}dX
+\half{}q\,d\overline{X}\wedge{}F\,dX
\label{JMS2form}
\end{equation}
where the \textit{dressed} mass may\footnote{Another proposal \cite{ChD}, coming from a \textit{dequantization} of the Dirac-Pauli equation, is
\begin{equation} 
M^2=m^2-\frac{gqs}{m}\,\alpha,
\label{ChD}
\end{equation} 
and coincides with (\ref{MJMS}) up to a higher-order term in the field
strength.} be given by
\begin{equation} 
M=m-\frac{g}{2}\frac{qs}{m}\,\alpha
\label{MJMS}
\end{equation} and where
\begin{equation}
\alpha=-\half\Tr(\Omega{}F)=\barI\star(F)J
\label{alpha}
\end{equation} is the spin-EMF \textit{coupling term}.
\goodbreak

The two-form~(\ref{JMS2form}) is indeed closed as a consequence of the homogeneous Maxwell equations, $dF=0$. One shows that it has generically rank $8$, by computing the rather complicated equations of motion, $\delta(X,I,J)\in\ker(\sigma_\mathrm{JMS})$, which turn out to be non-linear in~$F$; see, e.g., \cite{SSD,DFS,Kun,JMS,ChD}.

Now, in the special case of a constant electromagnetic field, \textit{linearization} of the above-mentioned equations around $F=0$ yields the one-dimensional distribution \cite{SSD}
\begin{equation}
\left\{
\begin{array}{rcl}
\delta{}X&=&\displaystyle
\lambda\left[I -
\frac{qs}{m^2}\left(1-\frac{g}{2}\right)\Omega{}FI\right]\\[8pt]
\delta{}I&=&\displaystyle
-\frac{q}{m}\lambda\,FI\\[8pt]
\delta{}J&=&\displaystyle
-\frac{q}{m}\lambda\left[\frac{g}{2}\,FJ + \left(1-\frac{g}{2}\right)I\,(\barI{}FJ)
\right]\\[8pt]
\end{array}
\right.
\label{JMS}
\end{equation}
on the evolution space, $V$, where $\lambda=\barI\delta{}X\in\bR$ (indicates a free parametrization of the integral leaves of this distribution). 
The system (\ref{JMS}) does exactly reproduce the two last BMT equations in
(\ref{BMT}) upon putting $P=m I$ for the linear momentum and $S=s J$ for the spin vector.
Let us emphasize that the velocity turns out to be no longer parallel to the linear momentum, due to the emergence of an \textit{anomalous velocity} (which vanishes in the \textit{normal} case, $g=2$). 

\section{A new model for non-minimal coupling: Stora's Ansatz}
\label{SectionRS}

\subsection{The Ansatz}

Let us start with a brand new, closed, two-form on evolution space $V$ (\ref{V}) which provides further modification to the presymplectic two-form (\ref{JMS2form}), viz.,
\begin{equation}
\medbox{
\hsigma
=
\half{}s\left(d\barI\wedge\Omega\,dI-d\barJ\wedge\Omega\,dJ\right)
-d\barP\wedge{}dX
+\half{}q\,d\overline{X}\wedge{}F\,dX
}
\label{RS2form}
\end{equation}
where, as suggested by Raymond Stora (RS), we \textit{posit} the new expression
\begin{equation}
\medbox{
P
=
(m+k\alpha)I+\ell\star(F)J
}
\label{P}
\end{equation}
of the ``linear momentum'' in (\ref{RS2form}), with $k,\ell\in\bR$ two independent parameters to be ultimately adjusted. This leads to a modification of the two-form $\sigma_\mathrm{JMS}^\mathrm{free}$ by terms merely linear in $F$. Notice, however, that we have to dispense with the usual (monolocality) constraint $\Omega{}P=0$.


We now endeavor to determine the parameters $k$ and $\ell$ by requiring that the null distribution of $\sigma$ duly lead again to the BMT equations.
Of course, one has
\begin{equation}
k=-\frac{g}{2}\frac{qs}{m}
\qquad
\mathrm{and}
\qquad
\ell=0
\label{kJMS}
\end{equation}
in Souriau's model (\ref{JMS2form}) and (\ref{MJMS}).

\goodbreak

The new equations of motion, $\delta(X,I,J)\in\ker(\hsigma)$, are thus
given by
\begin{equation}
\hsigma(\delta(X,I,J), \cdot\,)
+\lambda\,d(1-\barI{}I)+\mu\,d(1+\barJ{}J)+\nu\,d(\barI{}J)
=0
\label{kerBMT2form}
\end{equation}
where $\lambda,\mu,\nu$ are Lagrange multipliers taking care of
the constraints in~(\ref{V}). 

The general case of an arbitrary electromagnetic field is quite involved, and will not be required, at this point, to compare our distribution and the BMT differential equations. Therefore, in the approximation of a \textit{constant} electromagnetic field, $F$, we get from the linear system~(\ref{kerBMT2form}),
\begin{eqnarray}
\label{Eq1}
0&=&\displaystyle
-s\Omega\delta{}I+M\delta{}X+k(\barI{}\delta{}X)\star(F)J
-\lambda{}I+\nu{}J\\[8pt]
\label{Eq2}
0&=&\displaystyle
+M\delta{}I+k\,I\delta\alpha+qF\delta{}X
+\ell\star(F)\delta{}J\\[8pt]
\label{Eq3}
0&=&\displaystyle
+s\Omega\delta{}J-k(\barI{}\delta{}X)\star(F)I-\ell\star(F)\delta{X}+\mu{}J+\nu{}I
\end{eqnarray}
where we have put, this time,
\begin{equation}
M=m+k\alpha.
\label{M}
\end{equation}

\subsection{The linearized equations of motion}

Let us now linearize the system (\ref{Eq1})--(\ref{Eq3}) around the value $F=0$; in the
sequel, the notation ``$\approx$'' will stand for ``up to ${\cal O}(F^2)$ terms''.

\subsubsection{Determining the Lagrange multipliers}

The above Lagrange multipliers read as follows in this approximation, namely
\begin{eqnarray}
\label{nu}
\nu&\approx&0\\
\label{mu}
\mu&\approx&\alpha\left[k\barI{}\delta{}X+\frac{\ell\lambda}{m}\right]\\
\label{lambda}
\lambda&=&(m+2k\alpha)\barI{}\delta{}X.
\end{eqnarray}

We also find that Equation (\ref{Eq3}) yields
$\delta{}J\approx\widehat{\mu}\,FJ+\varrho{}I$ where $\widehat{\mu}=\mu/\alpha$
(see~(\ref{mu})) and 
$\varrho\approx-\barI{}FJ\left[k\barI{}\delta{}X+(\lambda/m)(\ell+qs/m)\right]$.
We furthermore obtain 
\begin{equation}
\delta\alpha\approx0.
\label{deltaalpha}
\end{equation}

The somewhat technical proof of Equations (\ref{nu})--(\ref{deltaalpha}) is deferred to Appendix~\ref{Proof1}.

\subsubsection{Linearizing the equations of motion}

With these preparations, we find
$$
\delta{}J
\approx
-\frac{q}{m}(\barI{}\delta{X})\left(-\frac{m}{qs}\right)\left[
(k+\ell)FJ-\left(k+\ell+\frac{qs}{m}\right)I(\barI{}FJ)
\right]
$$
and readily \textit{recover} the BMT equation (\ref{BMT}) for spin --- see also (\ref{JMS}). 

We indeed find
\begin{equation}
\delta{}J\approx-\frac{q}{m}(\barI{}\delta{}X)\left[\frac{g}{2}\,FJ +
\left(1-\frac{g}{2}\right)I\,(\barI{}FJ)
\right]
\label{deltaJ}
\end{equation}
provided
\begin{equation}
\medbox{
k+\ell=-\frac{g}{2}\,\frac{qs}{m}
}
\label{k+l}
\end{equation}
which is the \textit{unique} compatibility equation needed, at this stage,
for $k$ and $\ell$.

The Lorentz equation in (\ref{JMS}) is readily found to hold, namely
\begin{equation}
\delta{}I\approx-\frac{q}{m}(\barI{}\delta{}X)FI.
\label{deltaI}
\end{equation}

As for the velocity, it takes the following (provisional) form
\begin{equation}
\delta{}X\approx(\barI{}\delta{}X)\left[I-
\frac{1}{m}\left(k+\frac{qs}{m}\right)\Omega{}FI
\right]
\label{deltaX}
\end{equation}
where the normalized spin tensor $\Omega$ is as in (\ref{Omega}).

Note that the Souriau system is recovered from the distribution 
(\ref{deltaX}), (\ref{deltaI}), (\ref{deltaJ}), via the special values
(\ref{kJMS}), in accordance with (\ref{k+l}). Here, we gain
extra flexibility using the freedom provided by the single equation (\ref{k+l}) of compatibility for $k$ and $\ell$.

\subsubsection{Recovering the BMT equations}

The expression $\star(F)J=\Omega{}FI+\alpha{}I$ (see (\ref{twoUsefulFormulae})) is useful since it enables us to write, with the help of (\ref{k+l}), the the anomalous ``linear momentum'' (\ref{P}) as
$$
P=\left(m-\frac{g}{2}\,\frac{qs}{m}\,\alpha\right)I+\ell\,\Omega{}FI.
$$
Note that $P^2\approx{}m^2-(gqs/m)\alpha$, in accordance with (\ref{ChD}),
or\footnote{We assume that $P^2>0$ holds true in the weak field approximation, and put $\vert{}P\vert=\sqrt{P^2}$.}
$$
\vert{}P\vert\approx{}m-\frac{g}{2}\frac{qs}{m}\,\alpha.
$$

Let us now introduce the natural new unit vector \& normalized spin tensor\footnote{Note that $\Omega^*$ has the  Lorentz invariants of $\Omega$ in (\ref{Omega}), since 
$(I^*)^2=-J^2=1$ and $\barJ{}I^*=0$.}
\begin{equation}
I^*:=\frac{P}{\vert{}P\vert}
\qquad
\&
\qquad
\Omega^*:=j(I^*,J).
\label{NewIandOmega}
\end{equation}
The \textit{velocity} (\ref{deltaX}) then takes a new
form (derived in Appendix \ref{Proof2}), viz.,
\begin{equation}
\medbox{
\delta{}X\approx(\barI{}^*\delta{}X)\left[I^*
-\frac{qs}{m^2}\left(1-\frac{g}{2}\right)\Omega^*{}FI^*\right].
}
\label{NewdeltaX}
\end{equation}
With the above definitions, the \textit{Lorentz equation} is easily rewritten as
\begin{equation}
\medbox{
\delta{}I^*\approx-\frac{q}{m}(\barI^*{}\delta{}X)FI^*
}
\label{NewdeltaI}
\end{equation}
and the (last) BMT equation appears finally in its usual guise
\begin{equation}
\medbox{
\delta{}J\approx-\frac{q}{m}(\barI^*{}\delta{}X)\left[\frac{g}{2}\,FJ +
\left(1-\frac{g}{2}\right)I^*\,(\barI^*{}FJ)
\right].
}
\label{NewdeltaJ}
\end{equation}

We have therefore been able to recover the BMT equations with the single constraint (\ref{k+l}) These equations are therefore not sufficiently compelling to characterize the model, whence the further developments in the next section.

\section{The case of a static electromagnetic field \& spin-orbit coupling}\label{SO}

We have thus, at our disposal, a full-fledged presymplectic model $(V,\hsigma)$, yielding, in particular, equations of motion for spinning, charged, test particles endowed with an anomalous magnetic moment in an arbitrary (not necessarily constant) external electromagnetic field. Apart from providing, as seen above, the well-known BMT equations in the case of a weak constant  electromagnetic field, it will prove instrumental in our quest of a \textit{bona fide} spin-orbit coupling term in the case of an $(\SO(3)\times\bR)$-invariant, electric-like, field, $F$. This will constitute the decisive testing ground of the Ansatz (\ref{RS2form}) and (\ref{P}). 

The energy, $\cE$, and total angular momentum, $\vcJ$, of the system acquire in this formalism the unambiguous status of components of the associated moment map that we will now make explicit in order to fully complete the characterization of the parameters $k$ and $\ell$ defining the model.

\subsection{The $(\SO(3)\times\bR)$-moment map}\label{momentmapSection}

\subsubsection{Definition}

Let us recall the general definition of a moment map associated to a Lie group
action on a presymplectic manifold $(V,\sigma)$. 

Let $G$ be a Lie group (with Lie algebra $\gothg$) acting on $V$ and preserving the
closed two-form $\sigma$.  One says that $\Phi:V\to\gothg^*$ is a \textit{moment map} \cite{SSD} for these data if
\begin{equation}
\sigma(\xi_V,\cdot\,)=-d(\Phi(\xi))
\label{momentmap}
\end{equation} 
for all $\xi\in\gothg$, where $\xi_V$ denotes the fundamental vector field associated with $\xi$.

Note that $\Phi$ is constant along each leaf of the foliation $\ker(\sigma)$, thanks to~(\ref{momentmap}); it thus naturally factors through the space of classical states (or \textit{space of motions}) $V/\ker(\sigma)$; hence $\Phi$ is a Noetherian constant of the
motion.

\subsubsection{The moment map of time translations $(\bR,+)$}\label{energysubsubsection}

Let us assume that $F$ be stationary, i.e., invariant against
time-translations of space\-time whose action reads
\begin{equation}
X\mapsto{}X+eU
\qquad
\qquad
(e\in\bR)
\label{timetranslations}
\end{equation} 
where $U\in\bR^{3,1}$ is the (future-pointing) velocity of the observer
($U^2=1$). The natural lift of this $(\bR,+)$-action to $V$, given by
(\ref{timetranslations}) supplemented by $I\mapsto{}I$ and $J\mapsto{}J$, turns
out to trivially preserve the two form (\ref{RS2form}). 

Easy calculation (see Appendix \ref{Proof3} for details) leads to
the following expression of the \textit{energy}\footnote{Put $\Phi=-H$ and
$\xi\in\gothg\cong\bR$ in (\ref{momentmap}).}
\begin{equation}
H=\barU{}P+\phi
\label{Energy}
\end{equation} 
where $P$ is as in (\ref{P}) and $\phi$ is a scalar potential locally defined (up to an overall additive constant) by
\begin{equation}
F(U,\cdot\,)=d\phi.
\label{potential}
\end{equation} 

\subsubsection{The $\SO(3)$-moment map}


Rotations relatively to observer $U$, viz., 
$
\SO(3)\cong\{L\in\SO_+(3,1)\vert{}LU=U\}
$,
form a subgroup
of the (connected component of the) Lorentz group. We denote by $\imath_U:\so(3)\to\so(3,1)$ the corresponding homomorphism of Lie algebras and also by $\pi_U:\so(3,1)^*\to\so(3)^*$ the associated projection.

The natural infinitesimal action of rotations on the evolution space (given by the tangent lift of the action of rotations on space-time) reads
\begin{equation}
\delta(X,I,J)=(\Lambda{}X,\Lambda{}I,\Lambda{}J)
\qquad
\where\qquad
\Lambda\in\so(3,1)\ \&\ \Lambda{}U=0.
\label{rotationAction}
\end{equation} 
Since the electromagnetic field, $F$, is assumed to be also $\SO(3)$-invariant, this action turns out to be Hamiltonian, i.e., to yield a moment map
given by\footnote{\label{FootnoteJ} We use here the notation
$\Phi(\xi)=-\half\Tr(\cM\Lambda)=-\la\vcJ,\vomega\ra$ for all $\xi=\Lambda$ as in
(\ref{Lambda}) to define the angular momentum, $\vcJ$.}
\begin{equation}
\vcJ=\pi_U(\cM)
\qquad
\with
\qquad
\cM=X\barP -P\barX+s\Omega
\label{rotationMomenMap}
\end{equation} 
where $P$ has been defined in (\ref{P}) and $\Omega$ in (\ref{Omega}). A proof of
(\ref{rotationMomenMap}) is provided by Appendix \ref{Proof4}.

\subsubsection{Working in the Lab}

Let us introduce, here, the various components --- in a Lab frame --- of the physical
quantities we have previously introduced.

\goodbreak

The space-time event decomposes according to
\begin{equation}
X=\pmatrix{\vr\cr{}t}
\label{X}
\end{equation} 
with $\vr\in\bR^3$ and $t\in\bR$, whereas the observer unit velocity will be chosen as
\begin{equation}
U=\pmatrix{\v0\cr1}.
\label{U}
\end{equation} 

\goodbreak

The infinitesimal rotation generator (\ref{rotationAction}) is now\footnote{Here $j:\bR^3\to\so(3)$ is the Lie algebra isomorphism, $j(\vomega)\vr=\vomega\times\vr$, given by the cross product.}
\begin{equation}
\Lambda=\pmatrix{j(\vomega)&\v0\cr\v0&0}
\label{Lambda}
\end{equation} 
with $\vomega\in\bR^3$.

The vectors $I$ and $J$ decompose, accordingly, as\footnote{The brackets denote the
usual Euclidean scalar product of $\bR^3$.}
\begin{equation}
I=\gamma\pmatrix{\vv\cr{}1}
\qquad
\&
\qquad
J=\widetilde{\gamma}\pmatrix{\vu\cr\uv}
\label{IJ}
\end{equation} 
where $\vu,\vv\in\bR^3$ with $\Vert\vv\,\Vert<\Vert\vu\,\Vert=1$, while we have 
$\gamma=(1-\Vert\vv\,\Vert^2)^{-\half}$ \& $\widetilde{\gamma}=(1-\uv^2)^{-\half}$. 
This entails that 
\begin{equation}
\Omega
=
\gamma\widetilde{\gamma}\pmatrix{j(\vu-\vv\uv)&-\vu\times\vv\cr
-(\vu\times\vv)^T&0
}.
\label{OmegaLab}
\end{equation} 

We then introduce an electric field (satisfying $\star(F)U=0$) of the form
\begin{equation}
F
=
\pmatrix{0&\vE\cr
\vE^T&0
}
\qquad
\with
\qquad
\vE=-\phi'(r)\,\frac{\vr}{r}
\label{F}
\end{equation} 
where the potential $\phi$ is as in (\ref{potential}), and depends on
$r=\Vert\vr\,\Vert$ only. Let us record, for completeness, that
\begin{equation}
\star(F)
=
\pmatrix{-j(\vE)&\v0\cr
\v0&0
}.
\label{starF}
\end{equation} 
We readily get, from (\ref{OmegaLab}) and (\ref{F}), the following expression for the
spin-EMF coupling term, viz
\begin{equation}
\alpha
=
\gamma\widetilde{\gamma}\,\la\vE,\vu\times\vv\,\ra.
\label{alphaBis}
\end{equation} 

\subsection{The relativistic energy in the Lab}\label{EnergyLabSection}

\subsubsection{Energy}

Using the expression (\ref{IJ}) of the vectors $I$ and $J$, and the form
(\ref{starF}) of the Hodge dual of the field strength $F$, we find that the ''linear
momentum''~(\ref{P}) is, in the Lab, of the form
\begin{equation} 
P
=
\pmatrix{\vp\cr\cE}
=
\pmatrix{\gamma{}M\vv+\widetilde{\gamma}\ell\,\vu\times\vE\cr\gamma{}M}.
\label{PLab}
\end{equation} 
This implies that the energy (\ref{Energy}) relatively to $U$ is then
$H=\gamma{}M+\phi$, i.e.,
\begin{equation}
\medbox{
H=\gamma(m+k\alpha)+\phi
}
\label{EnergyLab}
\end{equation} 
where $\alpha$ is as in (\ref{alphaBis}). Note that
the extra coefficient $\ell$ does not show up here!

\subsubsection{Spin \& orbital momentum}

Wishing to express in the Lab frame the total angular momentum, $\vcJ$, found
in~(\ref{rotationMomenMap}), we compute
$$
\cM=\pmatrix{j(\vcJ)&\star\cr\star&0}
$$
where the ``$\star$'' are the boost-components we will not need to worry about. Using the expressions (\ref{X}), (\ref{PLab}) and (\ref{OmegaLab}), we get
$\vcJ=\vr\times\vp+s\gamma\widetilde{\gamma}\left(\vu-\vv\uv\right)$, and the
decomposition
\begin{equation}
\vcJ=\vL+\vS
\label{AngularMomentumLab}
\end{equation} 
where the \textit{orbital momentum} and the \textit{spin} read, respectively,
\begin{equation}
\medbox{
\vL=\vr\times\left(\gamma{}M\vv+\widetilde{\gamma}\ell\,\vu\times\vE\right)
\qquad
\&
\qquad
\vS=s\gamma\widetilde{\gamma}\left(\vu-\vv\uv\right).
}
\label{DecompositionAngularMomentumLab}
\end{equation}

\subsubsection{Decomposition of the energy}

Let us now compute the scalar product of the constituents of the angular momentum found in
(\ref{DecompositionAngularMomentumLab}); we find
\begin{eqnarray}
\la\vS,\vL\ra
&=&
s\gamma\widetilde{\gamma}\la\vu-\vv\uv,
\vr\times(\gamma{}M\vv+\widetilde{\gamma}\ell\,\vu\times\vE)
\ra\nonumber\\[6pt]
&=&
s\gamma^2\widetilde{\gamma}{}M\la\vu,\vr\times\vv\,\ra+\mathcal{O}(F)
\label{SL}
\end{eqnarray}
neglecting higher-order contributions in the field strength.

Equation (\ref{alphaBis}) giving the coupling spin-EMF term writes now, with the help of
(\ref{F}) and (\ref{SL}),
\begin{eqnarray*}
\alpha
&=&
\gamma\widetilde{\gamma}\,\frac{\phi'(r)}{r}\,\la\vu,\vr\times\vv\,\ra\\
&\approx&
\gamma\widetilde{\gamma}\,\frac{\phi'(r)}{r}\,\left(
\frac{1}{s\gamma^2\widetilde{\gamma}{}M}\,\la\vS,\vL\ra
\right)
\label{rotationMomentCalc}
\end{eqnarray*}
or, since $M=m+\mathcal{O}(F)$,
$$
\gamma\alpha
\approx
\frac{1}{sm}\,\frac{\phi'(r)}{r}\,\la\vS,\vL\ra.
$$

Returning to the expression (\ref{EnergyLab}) of the energy, we can claim that the decomposition
\begin{equation}
\medbox{
H\approx{}m\gamma+\frac{k}{sm}\frac{\phi'(r)}{r}\,\la\vS,\vL\ra+\phi
}
\label{goodEnergy}
\end{equation} 
helps us recover the expected coefficient in front of the spin-orbit coupling term 
provided
\begin{equation}
\medbox{
k=-\frac{(g-1)}{2}\,\frac{qs}{m}
}
\qquad
\&
\qquad
\medbox{
\ell=-\frac{1}{2}\,\frac{qs}{m}
}
\label{goodCoeffs}
\end{equation} 
in accordance with the BMT-constraint (\ref{k+l}). The expression (\ref{P}) of the linear momentum, $P$, is hence completely determined. 

\goodbreak

The remaining freedom in the choice of the parameters $k$ and $\ell$ has therefore been eliminated by the requirement that the spin-orbit term compare with the one
given by the Dirac equation in the case $g=2$; see \cite{Mes}. Equations (\ref{goodEnergy}) and~(\ref{goodCoeffs}) are also consistent with the formula (11.121) in \cite{Jac} that holds for any value of~$g$.

\section{Conclusion}\label{Conclusion}

We have first recalled the derivation of the BMT equations via Souriau's presymplectic model, spelled out in Section \ref{SectionJMS}, which is fairly well-accepted for describing non-minimal coupling of a relativistic spinning particle --- with anomalous magnetic moment --- to an arbitrary external electromagnetic field. Considering again this model, we have proved in Section \ref{SectionRS} that the Ansatz given by (\ref{RS2form}) \& (\ref{P}) bringing slight but crucial modifications helps us
\begin{itemize}
\item
recover the BMT equations (\ref{NewdeltaI})  and (\ref{NewdeltaJ}) in the weak field ap\-proxima\-tion, featuring an anomalous velocity (\ref{NewdeltaX}) already found in~\cite{JMS},
\item
restore, via Equation~(\ref{goodCoeffs}), the awaited correct expression of the spin-orbit coupling term  in the energy (\ref{goodEnergy}) associated with a static electric-like field.
\end{itemize}
The latter finding provides hence a reasonable solution to a subtle problem posed by the constraints (\ref{kJMS}) characterizing Souriau's model. However, as emphasized by Stora, Equation (\ref{goodCoeffs}) is clearly of a phenomenological nature, and should ideally arise unambiguously from first principles, which is not entirely the case as of today. One might think that a treatment \textit{\`a la} Kaluza-Klein could provide some useful intuition. In any case, pursuing the quest of a purely geometric framework for a \textit{predictive} model of classical particles with anomalous magnetic moment has been a program of research which we preciously kept in mind, and which, unfortunately, we have not had the opportunity to complete. 

As a closing personal remark, I would say that am glad to have this way contributed, with the late colleagues Raymond Stora and Valentine Telegdi, to advances in the understanding of the subtle classical description of spinning particles with anomalous magnetic moment in an external electromagnetic field.

\section*{Acknowledgement}

This work is dedicated to Raymond Stora, who discussed with me the length and breadth of the problem he raised some time ago, triggering these developments. I have strongly benefited from his expertise and from his kind and friendly support during the writing of the first draft of this article which owes much to him. I am also grateful to P\'eter Horv\'athy for his most constructive suggestions and encouragement. I finally wish to express my warmest thanks to Serge Lazzarini for his help, his advice, and his constant interest in this work. 


\section*{References}



\section{Appendix}\label{Appendix}

\subsection{Deriving Equations (\ref{nu})--(\ref{deltaalpha})}\label{Proof1}

\subsubsection{Vanishing of $\nu$}

Taking the scalar product of (\ref{Eq1}) and $J$, and using $\Omega{}J=0$, one finds
$\nu=M\,\barJ{}\delta{}X$. Again, the scalar product of (\ref{Eq3}) and $I$ yields
$\nu=\ell\,\barI\star(F)\delta{}X$.
Using (\ref{alpha}), and the useful
relationships
\begin{equation}
\star(F)I=\Omega{}FJ+\alpha{}J
\qquad\&\qquad
\star(F)J=\Omega{}FI+\alpha{}I,
\label{twoUsefulFormulae}
\end{equation}
we get
$\nu=-\ell(\barJ{}F\Omega\delta{}X+\alpha\,\barJ\delta{}X)
\approx
-\ell\alpha\,\barJ\delta{}X$
since $F\Omega\delta{}X\approx0$ because of (\ref{Eq1}) and~(\ref{Eq2}). Hence,
$\nu\approx-\ell\alpha\nu/M$ and thus $\nu(1+\ell\alpha/m)\approx0$, implying (\ref{nu}).

\subsubsection{Determination of $\mu$ and appearance of $\varrho$}

Evaluation of the scalar product of (\ref{Eq3}) and $J$ gives, using Definition (\ref{alpha}),
$\mu=k\alpha\,\barI\delta{}X-\ell\,\barJ\star(F)\delta{X}
\approx
k\alpha\,\barI\delta{}X-(\ell\lambda/m)\barJ\star(F)I$, and
\begin{equation}
\mu
\approx
\alpha\,\widehat{\mu}
\qquad
\with
\qquad
\widehat{\mu}
=
k\,\barI\delta{}X+\frac{\ell\lambda}{m}.
\label{muBis}
\end{equation}

Using (\ref{Eq3}), we find 
$
s\Omega\delta{}J
=
k\,\barI{}\delta{}X(\Omega{}FJ+\alpha{}J)+(\ell/M)\star(F)(\lambda{}I+{\cal O}(F))-\mu{}J
$,
since $\delta{}I={\cal O}(F)$, see (\ref{Eq2}).
Using (\ref{twoUsefulFormulae}), we get
$
\Omega\left(
s\delta{}J-(k\,\barI{}\delta{}X+\ell\lambda/m)\,FJ
\right)
\approx
\left[\alpha(k\,\barI{}\delta{}X+\ell\lambda/m)-\mu\right]J
\approx
0
$,
as clear from (\ref{muBis}). This tells us that
\begin{equation}
s\delta{}J
\approx
\varrho\,I+\widehat{\mu}\,FJ
\label{deltaJBis}
\end{equation}
with $\varrho\in\bR$ and $\widehat{\mu}$ as in (\ref{muBis}).

\subsubsection{Determining $\lambda$}

Taking the scalar product of (\ref{Eq1}) and $I$ leaves us with
$M\barI{}\delta{}X=\lambda-k\alpha\,\barI\delta{}X$. We therefore obtain
$
\lambda
=
(m+2k\alpha)\barI\delta{}X
$, and prove (\ref{lambda}).

\subsubsection{Determining $\varrho$}

The constraint $\barI{}J=0$ is preserved, which means that
$\delta\barI{}J+\barI\delta{}J=0$. This said, we get:
$
0=sM(\barI\delta{}J+\barJ\delta{}I)
=
M\barI(\varrho{}I+\widehat{\mu}\,FJ)+s\barJ(-qF\delta{}X-\ell\star(F)\delta{}J)
$,
using~(\ref{deltaJBis}) and (\ref{Eq2}). Thus, 
$\varrho{}M+\widehat{\mu}\,\barI{}FJ-(qs/M)\barJ{}F(\lambda{}I+{\cal
O}(F))+\varrho\ell\alpha\approx0$, and we can write, with the help of (\ref{M}) and
(\ref{muBis}),
\begin{equation}
\varrho
\approx
-\barI{}FJ\left(
k\,\barI\delta{}X+\frac{\lambda}{m}\left(\ell+\frac{qs}{m}\right)
\right).
\label{rho}
\end{equation}

\subsubsection{The term $\alpha$ is a constant of the motion}

Equations (\ref{deltaJBis}), (\ref{rho}), together with
$F\delta{}I\approx0$, imply
$\delta\alpha=\barI\star(F)\delta{}J\approx(\varrho/s)\barI\star(F)I=0$; hence
$\delta\alpha\approx0$, as in (\ref{deltaalpha}).

\subsection{Deriving Equation (\ref{NewdeltaX}) for the velocity}\label{Proof2}

The velocity (\ref{deltaX}) is written, using the definition (\ref{M}) for the
mass~$M$, as
\begin{eqnarray*}
\delta{}X&\approx&(\barI{}\delta{}X)\left[\frac{P}{M}-\frac{\ell}{m}(\Omega{}FI+\alpha{}I)
-\frac{1}{m}\left(k+\frac{qs}{m}\right)\frac{\Omega{}FP}{m}\right]\\
&\approx&(\barI{}\delta{}X)\left[\frac{P}{m}\left(1-\frac{k+\ell}{m}\,\alpha\right)
-\frac{1}{m}\left(k+\ell+\frac{qs}{m}\right)\frac{\Omega{}FP}{m}\right]\\
&\approx&(\barI{}\delta{}X)\left[\frac{P}{\vert{}P\vert}
-\frac{qs}{m^2}\left(1-\frac{g}{2}\right)\frac{\Omega{}FP}{\vert{}P\vert}\right]
\end{eqnarray*}
with the the constraint (\ref{k+l}). Use then the above definition for $I^*$ and
the fact that $I^*=I+{\cal O}(F)$ to justify that $\Omega^*=\Omega+{\cal O}(F)$
(see (\ref{NewIandOmega})). The awaited result, Equation (\ref{NewdeltaX}), then
follows.

\goodbreak

\subsection{Deriving the expression (\ref{Energy}) of the energy}\label{Proof3}

Let $(X,I,J)\mapsto\delta(X,I,J)=(\xi{}U,0,0)$ be the fundamental vector field of infinitesimal time-translations $\xi\in\bR$. Since $L_UF=0$ (where $L_U$
stands for the Lie derivative with respect to the vector field $U$), and $dF=0$, we
obtain $F(U,\cdot\,)=d\phi$ --- as in
(\ref{potential}) --- for some locally defined function $\phi$ (the scalar potential for $U$) of space-time. We also find $\delta\alpha=0$, and
return to (\ref{RS2form}) to write
$\sigma(\delta(X,I,J),\delta'(X,I,J))=\left[\barU\delta'{}P+\delta'\phi\right]\xi=\delta'\left[\barU{}P+\phi\right]\xi$, since $U$ is a constant vector field. We have just proved that
$$
\sigma(\delta(X,I,J),\,\cdot\,)=dH\,\xi
$$
where $H$, the $(\bR,+)$-moment map, is of the form (\ref{Energy}).

\goodbreak

\subsection{Deriving the expression (\ref{rotationMomenMap}) of angular momentum}%
\label{Proof4}

The vector field $X\mapsto\delta{}X=\Lambda{}X$ (where $\Lambda$ is as in (\ref{Lambda})) Lie-transports the electric-like field $F$; this just means that
$\delta{}F=(\partial{}F/\partial{}X)\delta{}X=[\Lambda,F]$.
We then find, using (\ref{alpha}) and (\ref{rotationAction}) that $\alpha$ is indeed
$\so(3)$-invariant, $\delta\alpha=0$. In view of~(\ref{P}), this readily entails that $\delta{}P=\Lambda{}P$. A useful result \cite{SSD} moreover states that
$\delta\barI\Omega\delta'{}I-\delta\barJ\Omega\delta'{}J
=
-\Tr(\delta\Omega\,\Omega\,\delta'\Omega)$.
Here we have $\delta\Omega=[\Lambda,\Omega]$.

The LHS of Definition (\ref{momentmap}) of the moment map is therefore written as
$$
\sigma(\delta(X,I,J),\delta'(X,I,J))
=
-s\Tr([\Lambda,\Omega]\,\Omega\,\delta'\Omega)
-\overline{\Lambda{}P}\delta'X+\delta'\barP\Lambda{}X
+q\,\overline{\Lambda{}X}F\delta'X.
$$

We now evaluate the RHS of the latter equation.
Since $\Omega^3=-\Omega$, we get
$
\Tr([\Lambda,\Omega]\,\Omega\,\delta'\Omega)
=
-\half\Tr(\Lambda\,\delta'\Omega)$.
We also have
$-\overline{\Lambda{}P}\delta'X+\delta'\barP\Lambda{}X=
\delta'\left(\barP\Lambda{}X\right)$. The last term actually vanishes because
$\overline{\Lambda{}X}F\delta'X=-\la\vomega,\vr\times\vE\ra\,\delta't$ using (\ref{X}), (\ref{Lambda}) and (\ref{F}); the electric field, $\vE$ being central, the result follows.
To sum up, we find that $\sigma(\delta(X,I,J),\delta'(X,I,J))=\half\Tr\left(
\delta'(X\barP -P\barX+s\Omega)\Lambda\right)$, hence
\begin{eqnarray*}
\sigma(\delta(X,I,J),\,\cdot\,)
&=&\half\,d\Tr\left(\left(X\barP -P\barX+s\Omega\right)\Lambda\right)\\
&=&\half{}d\Tr(\cM\Lambda)
\end{eqnarray*}
for all $\Lambda=\imath_U(\vomega)$ constrained to infinitesimal Euclidean rotations (\ref{rotationAction}) --- see footnote \ref{FootnoteJ}. This
establishes Formula (\ref{rotationMomenMap}) for the angular momentum.
 
\end{document}